\documentclass[english,10pt,aps,pra,twocolumn,groupedaddress,floatfix]{revtex4-1}
\pdfoutput=1
\usepackage[utf8]{inputenc}
\usepackage{amsmath}
\usepackage{amssymb}
\usepackage{amsfonts}
\usepackage{bm}
\usepackage{bbm}
\usepackage{epsfig}
\usepackage{grffile}
\usepackage{graphics}

\usepackage[usenames,dvipsnames]{color}
\definecolor{dblue}{rgb}{0,0.1,.6}

\usepackage[colorlinks=true,citecolor=dblue,linkcolor=dblue,urlcolor=dblue]{hyperref}
\usepackage[all]{hypcap}

\newcommand{\ud}{\mathrm{d}}
\newcommand{\Tr}{\operatorname{Tr}}
\newcommand{\bra}{\langle}
\newcommand{\ket}{\rangle}
\newcommand{\mc}[1]{\mathcal{#1}}
\newcommand{\pdag}{{\phantom{\dag}}}

\renewcommand{\H}{\mc{H}}
\newcommand{\hH}{\hat{H}}
\newcommand{\hN}{\hat{N}}
\newcommand{\hO}{\hat{O}}
\newcommand{\hc}{\hat{c}}
\newcommand{\vk}{\vec{k}}
\newcommand{\vx}{\vec{x}}
\newcommand{\vn}{\vec{n}}
\renewcommand{\vr}{\vec{r}}
\newcommand{\dm}{{\hat{\varrho}}}
\newcommand{\dms}{{\hat{\rho}}}
\renewcommand{\vec}[1]{{\boldsymbol{#1}}}
\newcommand{\A}{\mc{A}}
\newcommand{\B}{\mc{B}}
\newcommand{\N}{\mc{N}}
\renewcommand{\P}{\mc{P}}
\newcommand{\vol}{\operatorname{vol}}

\newcommand{\veps}{\varepsilon}
\newcommand{\gs}{\text{gs}}
\newcommand{\ETH}{\text{eth}}
\newcommand{\mce}{\text{mc}}
\newcommand{\gce}{\text{gc}}
\newcommand{\cft}{\text{cft}}
\newcommand{\therm}{\text{th}}
\newcommand{\eff}{\text{eff}}
\newcommand{\avg}{\text{avg}}

\newcommand{\duke} {Department of Physics, Duke University, Durham, North Carolina 27708, USA}

\newcommand{\Title} {Eigenstate entanglement: Crossover from the ground state to volume laws}
\newcommand{\Authors}
{
\author{Qiang Miao}
\affiliation{\duke}
\author{Thomas Barthel}
\affiliation{\duke}
}
\newcommand{\Date} {June 20, 2021}

\begin{document}

\title{\Title}
\Authors

\begin{abstract}
For the typical quantum many-body systems that obey the eigenstate thermalization hypothesis (ETH), we argue that the entanglement entropy of (almost) all energy eigenstates is described by a single crossover function. The ETH implies that the crossover functions can be deduced from subsystem entropies of thermal ensembles and have universal properties. These functions capture the full crossover from the groundstate entanglement regime at low energies and small subsystem size (area or log-area law) to the extensive volume-law regime at high energies or large subsystem size. For critical one-dimensional systems, a universal scaling function follows from conformal field theory (CFT) and can be adapted for nonlinear dispersions. We use it to also deduce the crossover scaling function for Fermi liquids in $d>1$ dimensions. The analytical results are complemented by numerics for large non-interacting systems of fermions in $d=1$, 2, and 3 dimensions and have also been confirmed for bosonic systems and non-integrable spin chains. Furthermore, one can deduce the distribution function for eigenstate entanglement.
\end{abstract}

\date{\Date}
\maketitle

\section{Introduction}\label{sec:intro}
A fundamental concept in modern physics and information theory is quantum entanglement. Specifically, the entanglement entropy quantifies quantum correlations and the utility of a given state for quantum information processing \cite{Bennett1996,Nielsen2000}. It is also used to guide tensor network state simulations and to bound their computation costs \cite{Verstraete2004-7,Vidal2006,Hastings2007-76}.
Henceforth, consider the entanglement entropy $S$ for the bipartition of a $d$-dimensional quantum many-body system into a compact subsystem $\A$ of volume $\ell^d$ and the rest $\B$, which is much larger or infinite. Ground states have been studied intensely \cite{Eisert2008,Latorre2009,Laflorencie2016-646} and one generally finds an area law, where $S$ is proportional to the surface \emph{area} of $\A$, or an area law with a logarithmic correction. In contrast, for random states and highly excited states, one generally finds $S$ to be proportional to the \emph{volume} of $\A$. The transition from the groundstate scaling to the extensive scaling and, more generally, the distribution of $S$ in excited states, have been largely unexplored.

The long-range physics of typical quantum many-body systems is captured by a field theory with local interactions.
Then the groundstate entanglement entropy in gapped systems obeys an area law $S_\gs\propto \ell^{d-1}$ \cite{Callan1994-333,Latorre2004,Calabrese2004,Plenio2005,Cramer2006-73,Hastings2007-76,Brandao2013-9,Cho2018-8,Kuwahara2020-11}. Intuitively, only the vicinity of the boundary between $\A$ and $\B$ contributes to $S_\gs$ due to the finite correlation length induced by the energy gap. In critical systems, the correlation length diverges and the scaling of $S_\gs$ depends on the dimension $d$ and particle statistics. Critical 1d systems are usually captured by CFT, giving $S_\gs\propto \ln \ell$ \cite{Srednicki1993,Callan1994-333,Holzhey1994-424,Vidal2003-7,Jin2004-116,Calabrese2004,Zhou2005-12}. For critical fermionic systems with a $(d-1)$-dimensional Fermi surface, $S_\gs$ generally obeys a log-area law \cite{Wolf2005,Gioev2005,Barthel2006-74,Li2006}
\begin{equation}\label{eq:Sgs}
	S_\gs\propto \ell^{d-1}\ln \ell.
\end{equation}
For critical bosonic systems in $d>1$ dimensions, $S_\gs$ still obeys the area law \cite{Srednicki1993,Callan1994-333,Barthel2006-74,Casini2009-42,Lai2013-111}.

Our previous understanding of excited states is rather limited. Area, log-area laws and subleading corrections were found for states with few-particle excitations (vanishing excitation-energy density) \cite{Das2006-73,Das2008-77,Masanes2009-80,Alcaraz2011-106,Berganza2012-01,Moelter2014-10,Moudgalya2018-98}
and for special rare excited states which are often ground states of other Hamiltonians \cite{Das2006-73,Alba2009-10,Moudgalya2018-98,Vafek2017-3}.
For broad classes of highly excited states, the entanglement volume law has been found in Refs.~\cite{Alba2009-10,Ares2014-47,Storms2014-89,Keating2015-338}.
Extensive scaling of the average eigenstate entanglement was shown in Refs.~\cite{Vdimar2017-119,Vidmar2017-119b,Vidmar2018-121,Lu2019-99,Huang2019-938}.

In this paper, we address the long-standing question about the scaling of $S$ in excited states and its transition from the groundstate scaling to an extensive scaling $S\propto\ell^d$ at higher energies. We argue and demonstrate that, generally, the eigenstate thermalization hypothesis (ETH) \cite{Deutsch1991-43,Srednicki1994-50,Rigol2008-452,Biroli2010-105,Beugeling2014-89,Kim2014-90,Alba2015-91,Lai2015-91,Dymarsky2018-97,Deutsch2018-81} implies the existence of crossover functions that capture the entanglement entropies of (almost) all eigenstates. Moreover, for system parameters and energies corresponding to the quantum critical regime of a system, the crossover function has universal scaling properties. For critical 1d systems, the result follows from CFT. We also derive the crossover scaling function for Fermi liquids. In addition, we discuss the scaling in gapped systems and the eigenstate entanglement distribution. The general arguments and derivations are confirmed numerically for systems in $d=1,2,3$ dimensions.

\section{ETH and excited-state entanglement}\label{sec:ETH}
According to the strong ETH, local expectation values of all energy eigenstates approach those of corresponding microcanonical ensembles with the same energy, where deviations decrease with increasing system size. Weak ETH allows for an exponentially small number of untypical energy eigenstates \cite{Biroli2010-105,Kim2014-90,Yoshizawa2018-120}. While the original notion of ETH concerns the convergence to thermal expectation values in dynamics, the essential hypothesis is in fact about the local features of the eigenstates. ETH is closely related to quantum typicality \cite{Popescu2006-2,Goldstein2006-96,Gemmer2004} which applies beyond energy constraints.
ETH implies that the entanglement entropies of excited states are basically given by the subsystem entropies of corresponding thermodynamic ensembles and that deviations vanish in the thermodynamical limit.

While strong ETH is difficult to establish in a general way, weak ETH \cite{Biroli2010-105,Mori2016_09,Iyoda2017-119} can be understood rather easily and, in contrast to strong ETH, also applies to integrable systems: Consider an observable $\hO$ with finite spatial support and the microcanonical ensemble $D^{-1}\sum_n |E_n\ket\bra E_n|$ for a small energy window $E-\Delta E\leq E_n\leq E$ containing $D$ energy eigenstates $|E_n\ket$. The weak ETH bounds the variance
\begin{equation}\label{eq:weakETH}
	\Delta O^2_\ETH:=D^{-1}\sum_n \big(\bra E_n|\hO|E_n\ket-\bra\hO\ket_\mce\big)^2
	 \leq \Delta O^2_\mce
\end{equation}
for deviations between eigenstate and microcanonical expectation values. The inequality, follows from $\bra E_n|\hO|E_n\ket^2\leq \bra E_n|\hO^2|E_n\ket$. For a translation-invariant system, $|E_n\ket$ can be chosen as momentum eigenstates and we can replace $\hO$ by the sum $\hO':=\frac{1}{\N}\sum_{i=1}^\N\hO_i$ over all lattice translates $\hO_i$ of $\hO$ without changing matrix elements in Eq.~\eqref{eq:weakETH}. This yields
\begin{equation}
	\Delta O^2_\ETH
	\leq \Delta O'^2_\mce = \frac{1}{\N^2} \sum_{i,j}\big(\bra\hO_i\hO_j\ket-\bra\hO_i\ket\bra\hO_j\ket\big)_\mce.
\end{equation}
Thus, if connected correlation functions decay exponentially or according to a sufficiently fast power law, $\Delta O_\ETH$ indeed vanishes in the thermodynamic limit,  $\lim_{\N\to\infty}\Delta O_\ETH=0$.
Because of the equivalence of thermodynamic ensembles for large systems \cite{Lebowitz1967-153,Ruelle1969,Mueller2015-340,Tasaki2018-172}, we can also use other canonical ensembles. We will employ the grand-canonical ensemble (GCE) $\dm_\gce$ with temperature and chemical potential chosen to match the energy and particle number of the energy eigenstates, i.e.,
\begin{equation}\label{eq:GCE}
	\dm_\gce=e^{-\beta(\hH-\mu\hN)}/Z\  \ \text{with}\, \ 
	\bra\hH\ket_\gce=E,\ \bra\hN\ket_\gce=N.
\end{equation}

The coincidence of eigenstate expectation values with thermal expectation values for all observables $\hO$ supported on $\A$, implies the coincidence of the corresponding subsystem density matrices, i.e.,
\begin{equation*}
	\bra E_n|\hO|E_n\ket\approx\bra\hO\ket_\gce \ \ \Rightarrow \
	\dms_n:=\Tr_\B |E_n\ket\bra E_n|\approx\Tr_\B \dm_\gce
\end{equation*}
for typical eigenstates $|E_n\ket$.
Hence, entanglement entropies $S_n(\ell)=-\Tr\dms_n\ln\dms_n$ of typical eigenstates are very close to subsystem entropies of the GCE and become extensive for large subsystems,
\begin{equation}\label{eq:crossover}
	S_n(\ell) \stackrel{\text{typical}}{\approx}
	S_\gce(\ell,\beta) \xrightarrow{\ell\gg\xi }\ell^d\, s_\therm(\beta),
\end{equation}
where $s_\therm(\beta)$ denotes the thermodynamic entropy density, and $\xi$ is the thermal correlation length. Equation~\eqref{eq:crossover} has important implications: (a) As long as (weak) ETH applies, the entanglement entropies of (almost) all eigenstates are captured by a single crossover function, determined by $S_\gce(\ell,\beta(E_n))$. (b) This function follows the groundstate entanglement scaling for small $\ell$ and crosses over to an extensive scaling at large $\ell$. (c) If the system parameters and energy (temperature) lie in a quantum critical regime of the considered model, general principles dictate that the entanglement entropies should follow a universal scaling function \cite{Shankar1994-66,Sachdev2011,Senthil2008-78}. 
Point (b) is due to a \emph{resolution limitation} effect: With observations on a subsystem $\A$ of linear size $\ell$, one cannot resolve variations of momentum-space Green's functions below a scale $\sim 1/\ell$. Hence, one can coarse-grain accordingly and, for $\ell$ below a crossover length $\ell_c$, the coarse-grained Green's functions of excited states approach that of the ground state. One recovers the extensive scaling predicted by thermodynamics for $\ell\gtrsim\ell_c$, and $\ell_c$ increases with decreasing energy ($\beta^{-1}$). More detail is provided in Appendix~\ref{sec:resolution}.

\section{Crossover in critical 1d systems}
\begin{figure}[t]
	\includegraphics[width=\columnwidth]{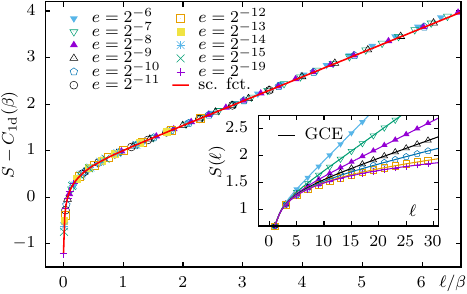}
	\caption{\label{fig:critical1d}	Entanglement entropies of randomly sampled eigenstates for a fermionic tight-binding chain with half filling ($\mu=0$, $v=2$), system size $L\approx 4\times 10^6$, excitation energy densities $e:=(E-E_\gs)/|E_\gs|$, and window size $\Delta E=1$. The main panel confirms the data collapse to the crossover scaling function \eqref{eq:ScftScaling}. The inset asserts the validity of the ETH by comparison to GCE subsystem entropies (lines).}
\end{figure}
Let us now investigate crossover functions for specific classes of systems. The long-range physics of critical 1d systems with linear dispersion at low energies, interacting or non-interacting, is described by 1+1d CFT. The GCE subsystem entropy can be computed using the replica trick and analytic continuation \cite{Korepin2004-92,Calabrese2004}. One obtains
\begin{equation}\label{eq:Scft}
	S^\cft_\gce(\ell,\beta) = \frac{c}{3}\ln\left[\frac{\beta v}{\pi a}\sinh\left(\frac{\pi\ell}{\beta v}\right)\right] + c'
\end{equation}
with the central charge $c$, group velocity $v$, ultraviolet cutoff $1/a$, and a nonuniversal constant $c'$. For small subsystem size $\ell$ or temperature $\beta^{-1}$, one recovers the log-area law $\frac{c}{3}\ln(\ell/a)$ [Eq.~\eqref{eq:Sgs}] as motivated above with the resolution argument. The crossover to extensive scaling $S\sim\frac{c}{3}\ell/\ell_c$ occurs at $\ell_c=\beta v/\pi$ and the universal scaling function is simply the leading term in
\begin{equation}\label{eq:ScftScaling}
	S^\cft_\gce(\ell,\beta)=\frac{c}{3}\ln(\sinh \ell/\ell_c)+\mc{O}(\ell^0).
\end{equation}
It applies whenever CFT does, including critical fermionic, bosonic and spin systems.

To confirm this numerically, we sample energy eigenstates $|E_n\ket$ from small windows of width $\Delta E$ around energies $E$ for fermionic tight-binding chains $\hH=-\sum_i(\hc_i^\dag\hc_{i+1}+h.c.)=\sum_k\veps_k\hat{n}_k$ at half filling. These obey weak ETH and are captured by a CFT with $c=1$. The sampling procedure is described in Appendix~\ref{sec:sampling}.
Figure~\ref{fig:critical1d} shows the results. The variances of the sampled $S_n(\ell)$ are much smaller than the symbol sizes. The inset asserts perfect agreement with the corresponding GCE subsystem entropies. The main plot shows the data collapse to the universal scaling function \eqref{eq:ScftScaling} after subtraction of $C_\text{1d}(\beta):=\frac{c}{3}\ln(\beta v/\pi a)$ and proper rescaling of $\ell$ with $\beta=\beta(E)$.

\section{Critical fermions in \texorpdfstring{$d>1$}{d\textgreater 1}}
\begin{figure}[t]
	\includegraphics[width=0.9\columnwidth]{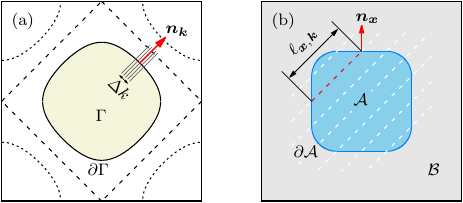}
	\caption{\label{fig:FermiSurface}(a) Fermi surfaces $\partial \Gamma$ for the 2d tight-binding model \eqref{eq:Htightbind} with fillings $\frac{1}{4}$, $\frac{1}{2}$, and $\frac{3}{4}$. (b) Real-space bipartition and chord lengths $\ell_{\vx,\vk}$ for direction $\vn_\vk$.}
\end{figure}

For higher-di\-men\-sio\-nal critical systems, it is more complex to extract crossover functions. Let us first discuss translation-invariant systems of non-interacting fermions with a $(d-1)$-dimensional Fermi surface and consider interactions later on. Employing the Widom conjecture \cite{Widom1982}, Gioev and Klich found the coefficient in the log-area law \eqref{eq:Sgs} for the groundstate entanglement as an integral over the Fermi surface $\partial\Gamma$ and the boundary $\partial \A$ of the subsystem \cite{Gioev2005,Leschke2014-112}. To leading order,
\begin{equation}\label{eq:SgsFermi}
	S_\gs(\ell)=\frac{\ln\ell}{12}\int_{\partial \A}\int_{\partial\Gamma}\frac{\ud A_x\ud A_k}{(2\pi)^{d-1}}\,|\vn_\vx\cdot\vn_\vk|
\end{equation}
with the normal vectors $\vn_\vx$ and $\vn_\vk$ on the boundary $\partial \A$ and on the Fermi surface $\partial \Gamma$ as indicated in Fig.~\ref{fig:FermiSurface}.
Eq.~\eqref{eq:SgsFermi} can be interpreted as an integral over entanglement contributions of lines perpendicular to the Fermi surface \cite{Swingle2010-105,Swingle2012-86}.

For the case of finite temperatures (energies), we can use the latter intuition and will involve the chord length as indicated in Fig.~\ref{fig:FermiSurface}b to correctly capture all temperature scales:
First consider the trivial example of uncoupled critical 1d chains oriented in a fixed direction $\vn_\vk$ with density $\rho_\perp$ in the perpendicular direction. To get the entropy $S_\gce(\ell,\beta)$ of a subsystem $\A$, we simply need to add the contributions from all chains. For convex $\A$ and the continuum limit, this gives $S_\A=\frac{\rho_\perp}{2}\int_{\partial \A}\ud A_x\,  |\vn_\vx\cdot\vn_\vk|\,S_\text{1d}(\ell_{\vx,\vk})$, where $\rho_\perp|\vn_\vx\cdot\vn_\vk|$ is the density of chains piercing $\partial \A$ at point $\vx$ and $\ell_{\vx,\vk}$ is the chord length across $\A$ in direction $\vn_\vk$ (see Fig.~\ref{fig:FermiSurface}b).
We now want to reduce the case of true $d$-dimensional systems with couplings in all directions to that of uncoupled chains as follows. Consider a patch $\P$ of size $\Delta k^{d-1}$ around $\vk\in\partial \Gamma$ on the Fermi surface and let us group wave vectors around that point into lines in direction $\vn_\vk$ as indicated in Fig.~\ref{fig:FermiSurface}a. We know that, at sufficiently low temperatures, modes far away from the Fermi surface are irrelevant for the long-range physics and can be disregarded. The dispersion is linear in direction $\vn_\vk$, corresponding to chiral fermions, but the dispersion is flat in the perpendicular directions. Let us parametrize these directions by $k_\parallel$ and $\vk_\perp$. Because of the flat dispersion with respect to $\vk_\perp$, any unitary transformation (basis change) among the single-particle states corresponding to $\vk_\perp\in\P$ does not generate any coupling of these modes. In particular, we can use this to transform to single-particle states that are spatially localized around points with spacings $\Delta y=2\pi/\Delta k$ in the $(d-1)$-dimensional perpendicular plane. A subsequent inverse Fourier transform with respect to $k_\parallel$ then yields uncoupled chains in direction $\vn_\vk$ and density $1/{\Delta y}$ in the directions perpendicular to $\vn_\vk$. This is the situation we considered initially with $\rho_\perp=1/\Delta y^{d-1}$, and we hence know how patch $\P$ contributes to the subsystem entropy. For $S_\text{1d}(\ell_{\vx,\vk})$, we can plug in the finite-temperature CFT result \eqref{eq:Scft}, where we substitute $\ell$ by the chord length $\ell_{\vx,\vk}$ and use $c=1/2$ because of chirality. Finally, we need to integrate over the entire Fermi surface to take into account the contributions from all patches, resulting to leading order in the subsystem entropy
\begin{equation}\label{eq:Shigher-d}
\begin{split}
	S_\gce(\ell,\beta)=\frac{1}{12}\int_{\partial \A}\int_{\partial\Gamma}&\frac{\ud A_x\ud A_k}{(2\pi)^{d-1}}\,|\vn_\vx\cdot\vn_\vk|\\
	&\times 	\ln\left[\frac{\beta v_\vk}{\pi a}\sinh\left(\frac{\pi\ell_{\vx,\vk}}{\beta v_\vk}\right)\right],
\end{split}
\end{equation}
where $v_\vk$ is the Fermi velocity at point $\vk\in\partial\Gamma$.
\begin{figure}[t]
	\includegraphics[width=\columnwidth]{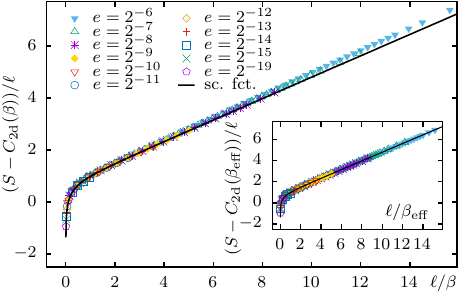}
	\caption{\label{fig:critical2dquarter} Eigenstate entanglement entropies for the critical 2d model \eqref{eq:Htightbind} with size $4096\times4096$ and $\mu\approx-1.44$ collapse onto a scaling function after subtraction of a subleading area-law term $C_\text{2d}(\beta)$ [Eq.~\eqref{eq:Cd}]. The latter corresponds to the factor $\beta/\pi a$ in the logarithm of Eq.~\eqref{eq:Shigher-d} with $\beta=\beta(E)$.}
\end{figure}

This formula for the subsystem entropy is remarkable: For zero temperature, we recover the groundstate result \eqref{eq:SgsFermi}. For large subsystems, the logarithm approaches $\pi\ell_{\vx,\vk}/\beta v_\vk$ such that the integral over $\partial \A$ gives the subsystem volume  $\vol\A$ and we are left with
\begin{equation}\label{eq:Sommerfeld}
	\frac{1}{6}\vol\A\int_{\partial\Gamma}\frac{\ud A_k}{(2\pi)^{d-1}}\frac{\pi}{\beta v_\vk}
	= \frac{\pi^2 g(\mu)}{3\beta}\vol\A.
\end{equation}
This is the well-known extensive thermodynamic entropy due to the Sommerfeld expansion \cite{Schwabl2006} with the density of states $g(\mu)$ at the Fermi energy.
Further, if we move the $\beta$ factor in the logarithm of Eq.~\eqref{eq:Shigher-d} to a subleading area law term and replace $\ell_{\vx,\vk}\equiv\ell\cdot\lambda_{\vx,\vk}$, where $\vol\A=\ell^d$ as before, the leading order term of $S/\ell^{d-1}$ is just a function of $\ell/\beta$. In this way, we obtain the desired scaling function for the crossover from groundstate to extensive subsystem entropies.

It is non-trivial to make the arguments leading to Eq.~\eqref{eq:Shigher-d} totally rigorous. The limit $\ell,\beta\to\infty$ has been captured using the theory of semiclassical trace formulas \cite{Widom1990-88,Sobolev2013-1043,Leschke2016-49,Leschke2017-273}. Equation~\eqref{eq:Shigher-d} should be treated as a conjecture. Our numerical tests in Fig.~\ref{fig:critical2dquarter} show, however, that it is very precise for all $\ell$ and $\beta$. The figure shows the data collapse of sampled eigenstate entanglement to the scaling function for fermions on a square lattice,
\begin{equation}\label{eq:Htightbind}
	\hH-\mu\hN=-\sum_{\bra i,j\ket}(\hc_i^\dag\hc_j^\pdag+h.c.)-\mu\sum_i\hc_i^\dag\hc_i^\pdag
\end{equation}
with $\mu\approx-1.44$ (quarter filling at zero temperature).
Deviations at large $\ell/\beta$ are due to the finite bandwidth of the model, i.e., due to a nonlinear dispersion at higher energies, differing from the assumptions of CFT. The inset shows that this can be fixed by replacing $\beta$ in Eq.~\eqref{eq:Shigher-d} by $\beta_\eff$. Specifically, we define an effective temperature by matching the exact thermodynamic entropy density and the large-$\ell$ limit \eqref{eq:Sommerfeld} of Eq.~\eqref{eq:Shigher-d}, i.e., $s_\therm(\beta)=:{\pi^2 g(\mu)}/{3\beta_\eff(\beta)}$ such that $\beta_\eff(\beta)\to\beta$ at low temperatures. The substitution guarantees that Eq.~\eqref{eq:Shigher-d} reproduces the correct thermodynamic entropy density at large $\ell/\beta$ and, in a sense, straightens the dispersion relation.
If $v_\vk$ is zero at some points on the Fermi surface, CFT and Eq.~\eqref{eq:Shigher-d} are not applicable. But the described rescaling procedure still works as we demonstrate in Appendix~\ref{sec:2dHalfFill} for the 2d tightbinding model \eqref{eq:Htightbind} at half filling. Equation~\eqref{eq:Shigher-d} works without issues for 3d as shown in Appendix~\ref{sec:3dHalfFill}.

The crossover scaling function \eqref{eq:Shigher-d} also applies to \emph{interacting} metals described by Fermi liquid theory, because the quasi-particle lifetime diverges when approaching the Fermi energy. The scaling function is universal in the sense that the only remnant of microscopic details is the dependence on the Fermi surface shape and $v_\vk$ as pointed out for other quantities in Ref.~\cite{Senthil2008-78}.

\section{Crossover in gapped systems}
If one adds a mass term to a critical theory, the resulting gap $\sim m$ represents an additional energy scale. At zero temperature, the entanglement entropy should then be determined by a function of $m\ell$ and this has indeed been confirmed for several cases \cite{Casini2005-07,Casini2005-12,Casini2009-42,Cardy2008-130,CastroAlvaredo2008-41,Doyon2009-102}. So far, very few works investigated excited-state entanglement entropies in gapped systems \cite{Storms2014-89} and, to our knowledge, the crossover from area law to extensive scaling and its universal properties have not been addressed.

When the total system gets sufficiently large compared to the considered subsystem, the entanglement entropies of (almost) all energy eigenstates converge to the subsystem entropy of the corresponding GCE due to ETH. These subsystem entropies are, in principle, functions of subsystem size $\ell$, mass $m$, and $\beta$. However, we expect that it can be expressed in terms of a scaling function that only depends on the parameters $m\ell$ and $m\beta$, characterizing the full crossover behavior. For fermionic systems, in particular, it follows a log-area law as in Eq.~\eqref{eq:Sgs} for $m\ell\ll \min(1,m\beta)$ and a volume law for $m\ell\gg \min(1,m\beta)$. Moreover, at low temperatures, $m\beta\gg 1$, one has the typical behavior of thermally activated excitations, and the growth of the entropy density $s_\therm$ in Eq.~\eqref{eq:crossover} as a function of temperature changes from exponential to linear around $m\beta\sim 1$. As an example consider massive Dirac fermions, i.e., the energy-momentum relation $\veps_\vk=\sqrt{m^2+(v\vk)^2}$.
The scaling form is most easily exemplified for the volume-law regime: The thermodynamic entropy is
\begin{equation*}
	S_\therm=-(\ell/2\pi)^d\int\ud^d k\left[f_\vk\ln f_\vk+(1-f_\vk)\ln(1-f_\vk)\right]
\end{equation*}
with the Fermi-Dirac distribution $f_\vk=1/\left(1+e^{\beta\veps_\vk}\right)$. Substituting $\vec{q}:=v\vk/m$, the integral becomes a function of $m\beta$ only with prefactor $m\ell$ as predicted.
\begin{figure}[t]
	\includegraphics[width=\columnwidth]{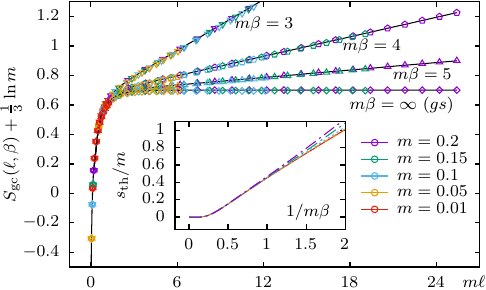}
	\caption{\label{fig:gapped1d}
	Subsystem entropies $S_\gce(\ell,\beta)$ in the staggered tight-binding chain \eqref{eq:Hstaggered} with half filling ($\mu=0$), mass $m$, and $L\approx 10^6$. We only use odd subsystem sizes $\ell$ because of odd-even effects. Here, lines are just guides to the eye.}
\end{figure}

Figure~\ref{fig:gapped1d} shows the data collapse of GCE subsystem entropies to the two-parameter scaling function for the staggered tight-binding chain
\begin{equation}\label{eq:Hstaggered}
	\hH=-\sum_{i}\left(t\hc_{2i-1}^\dag\hc_{2i}^\pdag+t'\hc_{2i}^\dag\hc_{2i+1}^\pdag+h.c.\right)
\end{equation}
at half filling. The dispersion relation is $\veps_k=\pm\sqrt{t^2+t'^2+2tt'\cos k}$ with mass $m=|t-t'|$ and $v=\sqrt{tt'}$. It takes the relativistic form $\pm\sqrt{m^2+(vp)^2}$ for small $p=k\pm\pi$. The inset displays scaled entropy densities $s_\therm$ which show the thermally activated exponential behavior at low temperatures $m\beta\gg 1$ and linear scaling for higher temperature.

\section{Entanglement distribution}
Recently, interesting results were derived for the entanglement entropy \emph{averaged over all} energy eigenstates \cite{Vdimar2017-119,Vidmar2017-119b,Vidmar2018-121,Lu2019-99,Huang2019-938}. In particular, bounds on the average entanglement for chaotic local Hamiltonians and quadratic fermionic systems were given. As long as one only considers the case of small subsystems and the (weak) ETH is applicable, which holds quite generally as discussed above, we can get much more -- the \emph{entire distribution} of eigenstate entanglement.
We have shown that the eigenstate entanglement entropies are given by certain crossover functions. To obtain the entanglement distribution, we only need to multiply these with the many-body density of states. For many purposes, the latter can be approximated by a Gaussian that describes well the bulk of the spectrum \footnote{The approach to a Gaussian in the bulk of the spectrum follows from the central limit theorem for non-interacting systems and can be derived more generally with random matrix theory \cite{French1970-33,Brody1981-53}. At the spectral edges, the density of states is exponentially suppressed and follows, e.g., power laws \cite{Hastings2007-76} or stretched exponentials \cite{Bethe1936-50,Cardy1986-270,Verlinde2000_08}. These do not contribute in the thermodynamic limit.}.

From this, properties of the average entanglement follow rather easily. For example, Ref.~\cite{Vdimar2017-119} addresses for non-interacting translation-invariant fermions how the average entanglement between a subsystem of fixed size $\ell^d$ converges to the maximum $\bra S\ket=\ell^d$ in the thermodynamic limit. This can also be explained as follows. With the same arguments as in the derivation for the weak ETH in Sec.~\ref{sec:ETH},
we can bound the deviation of expectation values $\bra E_n|\hO|E_n\ket$ of a local observable from the infinite temperature value $\bra\hO\ket_\infty=\Tr(\hO)/D$, averaged over \emph{all} energy eigenstates, by 
$\Delta O^2_\avg:=D^{-1}\sum_{n=1}^D \big(\bra E_n|\hO|E_n\ket-\bra\hO\ket_\infty\big)^2 \leq \Delta O^2_\infty$, where $D=\dim\H$. And $\Delta O_\infty$ decays with $\mc{O}(\N^{-1/2})$ in the thermodynamic limit $\N\to\infty$. Hence, almost all energy eigenstates
look locally like the infinite temperature ensemble and their entanglement entropies are maximal.

\section{Discussion}
In conclusion, ETH can be employed to understand the full crossover of eigenstate entanglement entropies from the groundstate scaling at small subsystem sizes and low energies to the extensive scaling at large sizes and higher energies. With increasing system size, the entanglement entropies of all (strong ETH) or almost all (weak ETH) energy eigenstates converge onto a single crossover function, which can be determined from corresponding subsystem entropy of corresponding thermodynamic ensembles -- also in integrable systems. Importantly, for the quantum critical regime, one obtains \emph{universal} scaling functions that capture large classes of microscopic models \cite{Shankar1994-66,Sachdev2011,Senthil2008-78}. For critical 1d systems and fermions in $d>1$ dimensions, analytic forms for the scaling functions were given and numerically confirmed. We substantiate the scaling properties further for bosonic systems in Ref.~\cite{Barthel2019_12} and for interacting integrable and non-integrable spin chains in Ref.~\cite{Miao2020_10}.
These results generalize immediately to R\'{e}nyi entanglement entropies and can, for example, be used to derive upper bounds on computation costs in tensor network simulations \cite{Verstraete2005-5,Barthel2017_08}.

Scaling functions for thermal subsystem entropies are so far largely unexplored. The connection to eigenstate entanglement makes them very interesting and it will be an exciting endeavor to derive crossover functions for specific lattice models and field theories, similar to efforts on groundstate entanglement. Furthermore, the accuracy of the $d>1$ scaling function \eqref{eq:Shigher-d} for fermions suggests an extension of the famous Widom formula for groundstate entanglement to finite temperatures.

\begin{acknowledgments}
We gratefully acknowledge helpful discussions with Giulio Biroli, Pasquale Calabrese, Anatoly Dymarsky, Takashi Ishii, Israel Klich, Jianfeng Lu, Marcos Rigol, Takahiro Sagawa, and Xin Zhang
and support through US Department of Energy grant DE-SC0019449.
\end{acknowledgments}

\appendix

\section{Sampling for non-interacting systems}\label{sec:sampling}
In the simulations for non-interacting fermionic systems, we sample energy eigenstates $|\{n_\vk\}\ket$ from windows of width $\Delta E\sim 1$ around a given energy $E$. In each update, occupation numbers $n_\vk$ and $n_{\vk'}$ for two randomly chosen wave vectors $\vk$ and $\vk'$ are swapped if the energy stays in the predefined window, In this way, the particle number $N=\sum_\vk n_\vk$ stays constant. The eigenstates are Gaussian states and, according to Wick's theorem \cite{Fetter1971}, fully characterized by the single-particle Green's function
\begin{equation}\label{eq:GF1x}
	G_{i,j}:=\bra \hc_i^\dag\hc_j^\pdag\ket.
\end{equation}
The entanglement entropy is computed through a diagonalization of $G_{i,j}$ restricted to sites $i$ and $j$ of the considered subsystem $\A$. From the eigenvalues $\nu_q$, one obtains the entanglement entropy in the form
\begin{equation}
	S_\A=-\sum_q\left[\nu_q\ln \nu_q+(1-\nu_q)\ln(1-\nu_q)\right].
\end{equation}

\section{Resolution limitation}\label{sec:resolution}
In the main text, a resolution limitation effect was mentioned that gives an alternative motivation for the ETH and explains why the entanglement crossover functions transition to the groundstate scaling for small subsystem sizes with $\ell<\ell_c$.

It can be understood most easily for non-interacting translation-invariant fermionic systems. In this case, energy eigenstates are characterized by occupation numbers $\{n_\vk\}$ for Bloch states. These determine the single-particle Green's function \eqref{eq:GF1x} according to
\begin{equation}\label{eq:GF1k}
	G_{i,j} = \frac{1}{L^d}\sum_\vk n_\vk e^{-i\vk\cdot(\vr_i-\vr_j)},
\end{equation}
where $\vr_i$ is the position of site $i$ and the total number of sites is $\N=L^d$. As already mentioned in Appendix~\ref{sec:sampling}, Wick's theorem \cite{Fetter1971} allows us to express the expectation value of any observable which is supported on subsystem $\A$ (and the identity in its complement $\B$) through $G_{i,j}$ with $i,j\in\A$. Let $\ell:=\max_{i,j\in A}|\vr_i-\vr_j|$ be the diameter of $\A$. Then, observations on $\A$ cannot resolve variations of $n_\vk$ on scales below $\sim 1/\ell$. Specifically, we can define coarse-grained variables through the convolution
\begin{equation}
	\tilde{n}_\vk:=\frac{1}{L^d}\sum_{\vk'} g(\vk-\vk')n_{\vk'}
\end{equation}
with a filter function $g$, say a Gaussian with standard deviation $\sigma_k$. Expanding the exponential in Eq.~\eqref{eq:GF1k} shows that replacing $n_\vk\to\tilde{n}_\vk$ in Eq.~\eqref{eq:GF1k} changes $G_{i,j}$ on the order of $\ell \sigma_k$. This deviation in $G_{i,j}$ vanishes for $\sigma_k\ll 1/\ell$ which has two important implications.
\begin{figure}[t]
	\includegraphics[width=0.98\columnwidth]{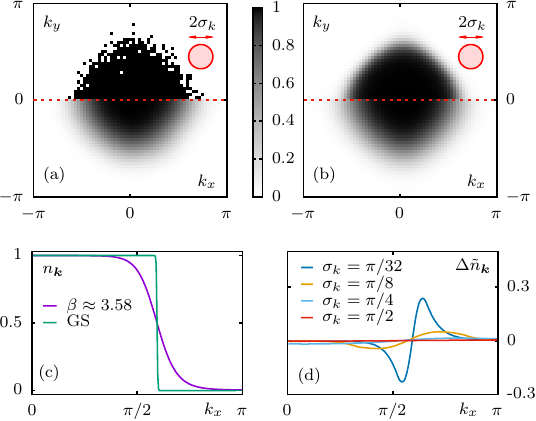}
	\caption{\label{fig:resolution} Illustration of the resolution-limitation effect.
	(a) Occupation numbers $n_\vk\in\{0,1\}$ for a random eigenstate of the fermionic 2d tight-binding model at quarter filling, system size $\N=64\times 64$, and excitation energy density $e=1/64$ (shown for $k_y>0$). The coarse-grained distribution $\tilde{n}_\vk$ with $\sigma_k = \pi/8$ is shown for $k_y<0$.
	(b) Fermi-Dirac distribution for the same energy density shown for $k_y>0$ and its coarse-grained version shown for $k_y<0$. The latter is indistinguishable from $\tilde{n}_\vk$ of panel (a) in agreement with the ETH.
	(c) Groundstate distribution and the Fermi-Dirac distribution for $e=1/64$ ($\beta\approx 3.58$, $\mu\approx -1.44$); both with quarter filling, system size $\N=512\times 512$, and $k_y=0$.
	(d) Differences between coarse-grained versions of the two distributions of panel (c) for different standard deviations $\sigma_k$ of the Gaussian filter. They become indistinguishable for $\sigma_k\gtrsim \pi/\beta v$.}
\end{figure}

(a) A variation of the $\tilde{n}_\vk$ according to the maximum entropy principle \cite{Jaynes1957-106,Lai2015-91} shows that their most probable value is given by the convolution of the Fermi-Dirac distribution
\begin{equation}\label{eq:FD}
	\bra \hat{n}_\vk\ket_\gce = 1/\big(e^{\beta(\veps_\vk-\mu)}+1\big)
\end{equation}
with the filter function $g$.
The main text contains a very general proof of weak ETH based on a sufficiently fast decay of connected spatial correlation functions. Accordingly, a sufficiently smooth dispersion relation $\veps_\vk$ results in almost all eigenstates having $\{\tilde{n}_\vk\}$ in the immediate vicinity of the filtered Fermi-Dirac distribution which implies weak ETH.

(b) The ability to coarse grain $n_\vk$ on a scale $\sigma_k\ll 1/\ell$ also means that eigenstates below a certain energy density become locally indistinguishable from the ground state, which explains why the entanglement crossover functions transition to the groundstate scaling for small subsystem sizes with $\ell\ll\ell_c$. Specifically, this happens when $\ell \ll \beta v/\pi$, where $\beta=\beta(E)$ is the inverse temperature for the considered eigenstate energy, and $v$ is the Fermi velocity: The Fermi-Dirac distribution \eqref{eq:FD} goes from $1$ to $0$ in a momentum range $\sim \pi/\beta v$ around the Fermi energy. If we choose $\sigma_k\gtrsim \pi/\beta v$ for the Gaussian filter, the filtered thermal Fermi-Dirac distribution becomes indistinguishable from the filtered groundstate distribution. This is consistent with the crossover length $\ell_c=\beta v/\pi$ found from CFT.
\begin{figure*}[t]
	\includegraphics[width=\textwidth]{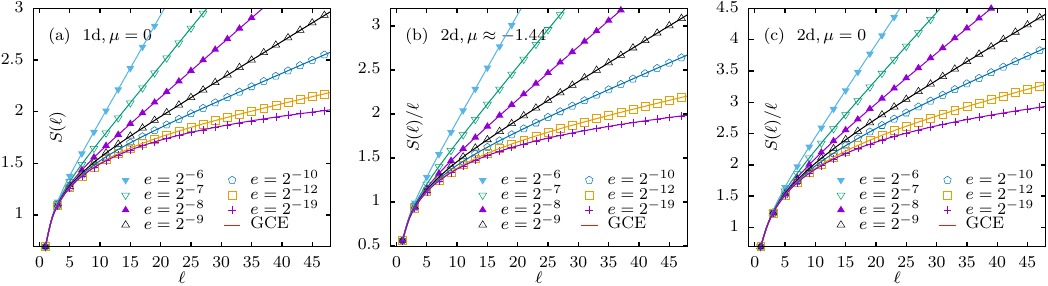}
	\caption{\label{fig:confirmETH} Confirmation that ETH is applicable for the description of eigenstate entanglement entropies in the fermionic tight-binding models. Points are averaged entanglement entropies for randomly sampled eigenstates in energy windows of size $\Delta E=1$ with excitation energy density $e$. Standard deviations are smaller than the symbol sizes. Lines show the subsystem entropy in the corresponding GCE with $\bra\hH\ket_\gce=E$ and chemical potential $\mu$. (a) 1d system at half filling ($\mu=0$) and size $L\approx 4\times 10^6$. (b) 2d square lattice with quarter filling at zero temperature ($\mu\approx-1.44$) and size $4096\times4096$. (c) 2d square lattice with half filling ($\mu=0$) and size $4096\times4096$.}
\end{figure*}

These arguments apply similarly for interacting systems. The major difference is that Wick's theorem does not apply anymore. Hence, one needs to consider multi-particle Green's functions. Expectation values for observables supported on $\A$ can be expressed by Fourier transforms of $n$-particle momentum-space Green's functions. Similarly to the arguments above, we can coarse-grain their momentum dependencies in accordance with the linear subsystem size $\ell$. For sufficiently small $\ell$, the coarse-grained Green's functions of eigenstates and corresponding thermal ensembles become indistinguishable from the coarse-grained groundstate Green's functions.

Things are very similar for bosonic systems. In the non-interacting case, one difference to fermions is that bosonic energy eigenstates are not Gaussian. So one needs to work right away with $n$-particle Green's functions or one can mimic eigenstates with squeezed states as discussed in Ref.~\cite{Barthel2019_12}. The ground state and thermal equilibrium states of non-interacting bosonic systems are still Gaussian.
\begin{table}[b]
\setlength{\tabcolsep}{5.1pt}
\begin{tabular}{l|l|lll|ll}
      &\multicolumn{1}{c|}{1d, $\mu=0$}&\multicolumn{3}{c|}{2d, $\mu\approx-1.44$}&\multicolumn{2}{c}{2d, $\mu=0$}\\
      &\multicolumn{1}{c|}{\small$\N\!\approx\! 4\!\cdot\! 10^6$}&\multicolumn{3}{c|}{\small$\N=4096^2$}&\multicolumn{2}{c}{\small$\N=4096^2$}\\
  \hline
  $e$ & $\beta$ & $\beta$ & $\beta_\eff/\beta$ & $N/\N$ & $\beta$ & $\beta_\eff/\beta$\\
  \hline
$2^{-6}$	&5.228	&3.583	&0.956	&0.2547	&5.034 	&2.12\\
$2^{-7}$	&7.319	&5.021	&0.978	&0.2523	&7.472 	&1.95\\
$2^{-8}$	&10.30	&7.066	&0.990	&0.2511	&11.04 	&1.80\\
$2^{-9}$	&14.54	&9.962	&0.995	&0.2506	&16.25 	&1.67\\
$2^{-10}$	&20.54	&14.05	&0.998	&0.2503	&23.84 	&1.56\\
$2^{-11}$	&29.03	&19.79	&0.999	&0.2501	&34.88 	&1.47\\
$2^{-12}$	&41.07	&27.75	&0.999	&0.2501	&50.94 	&1.39\\
$2^{-13}$	&58.11	&38.74	&1.000	&0.2500	&74.13 	&1.31\\
$2^{-14}$	&81.96	&53.27	&1.000	&0.2500	&108.0 	&1.25\\
$2^{-15}$	&116.2	&71.53	&1.000	&0.2500	&156.5 	&1.19\\
$2^{-19}$	&476.8	&143.1	&1	&0.2500	&667.6 	&1
\end{tabular}
\caption{\label{tab:beta}Energies $E$, given in terms of the excitation energy density $e$, and corresponding inverse GCE temperatures $\beta=\beta(E)$ in the fermionic 1d and 2d tight-binding models with chemical potential $\mu$. For the 2d system with $\mu\approx-1.44$, chosen to have quarter-filling at zero temperature, we also specify the corresponding filling factors $N/\N$.}
\end{table}

\section{Assertion of the ETH for critical fermions}\label{sec:assertETH}
Figure~\ref{fig:confirmETH} confirms the applicability of the ETH for the description of eigenstate entanglement entropies in the fermionic tight-binding models in one and two dimensions with $\N$ lattice sites. Points are averaged entanglement entropies for randomly sampled eigenstates in energy windows of size $\Delta E=1$ with excitation energy density $e=(E-E_\gs)/|E_\gs|$.

For each energy $E$, the sampling was done for $~16000\N$ iterations, and entanglement entropies were computed every $16\N$ iterations. The results are well converged. Standard deviations are smaller than the symbol sizes. Lines in the figure show the subsystem entropy in the corresponding GCE with $\beta=\beta(E)$ such that $\bra\hH\ket_\gce=E$, and chemical potential $\mu$. Particle numbers for the eigenstates were chosen to match the GCE expectation value $N=\bra\hN\ket_\gce$. Because of the particle-hole symmetry, $N=\N/2$ for all $\beta$ at $\mu=0$. The figure asserts agreement of the eigenstate entanglement with the GCE subsystem entropies.

Table~\ref{tab:beta} shows the relation between excitation energy densities and temperatures of the corresponding GCE for the fermionic tight-binding models in 1d and 2d. For 2d with quarter filling at zero temperature ($\mu\approx -1.44$), the table also shows the temperature-dependent filling factor. For the 2d systems, the effective temperatures, introduced to compensate for the nonlinearity of the dispersion relation, are given as well. They converge to the ``bare'' temperature $\beta^{-1}$ in the low-energy regime.

\section{Entanglement crossover for 2d fermions at half filling}\label{sec:2dHalfFill}
\begin{figure}[t]
	\includegraphics[width=0.98\columnwidth]{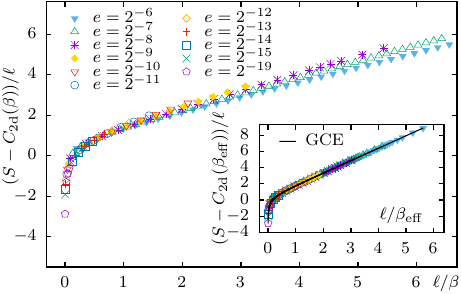}
	\caption{\label{fig:critical2dHalf} Eigenstate entanglement entropies for the critical 2d model \eqref{eq:Htightbind} at half filling ($\mu=0$) with system size $4096\times4096$, and various excitation energy densities $e$. Deviations from the scaling function at large $\ell/\beta$ are due to the finite bandwidth and can be removed by using instead the effective temperature $\beta^{-1}_\eff$ as shown in the inset.}
\end{figure}
The Fermi surface $\partial\Gamma$ for the critical 2d tight-binding model \eqref{eq:Htightbind} at half filling is shown in Fig.~\ref{fig:FermiSurface}a. In this case, the Fermi velocity $v_\vk$ vanishes at the points $\vk=(0,\pm\pi)$ and $(\pm\pi,0)$ where the Fermi surface touches the Brillouin zone boundary. Consequently, CFT and Eq.~\eqref{eq:Shigher-d} are not applicable. But the described rescaling procedure works nevertheless as shown in Fig.~\ref{fig:critical2dHalf}. As for the case without vanishing $v_\vk$, we subtract from the entanglement entropy $S_\A$ of the sampled energy-$E$ eigenstates the subleading area term
\begin{equation}\label{eq:Cd}
	C_{d>1}(\beta) = \frac{1}{12}\int_{\partial \A}\int_{\partial\Gamma}\frac{\ud A_x\ud A_k}{(2\pi)^{d-1}}\,|\vn_\vx\cdot\vn_\vk|\,
	\ln({\beta}/{\pi a}),
\end{equation}
corresponding to the factor $\beta/\pi a$ in the logarithm of Eq.~\eqref{eq:Shigher-d}, and we plot the result as functions of $\ell/\beta$ with $\beta=\beta(E)$. The data for different energies $E$, clearly collapses to a scaling function.

Deviations from the scaling function at larger $\ell/\beta$ are due to the nonlinearity of the dispersion. They can again be removed by using $C_\text{2d}(\beta_\eff)$ instead of $C_\text{2d}(\beta)$ and plotting against $\ell/\beta_\eff$ instead of $\ell/\beta$ as shown in the inset of Fig.~\ref{fig:critical2dHalf}. At low temperatures, the thermodynamic entropy density $s_\therm$ is linear in the temperature $\beta^{-1}$. We use that regime to determine $\sigma_\therm$ in $s_\therm=\sigma_\therm/\beta +\mc{O}(\beta^{-2})$ and define the effective temperature $\beta_\eff^{-1}$ through
\begin{equation}
	s_\therm(\beta)=:\sigma_\therm/\beta_\eff(\beta)
\end{equation}
for all temperature scales. As before, $\beta_\eff(\beta)\to\beta$ at low temperatures. See Table~\ref{tab:beta} for the relation between $E$, $\beta$, and $\beta_\eff$.
\begin{figure}[b]
	\includegraphics[width=0.98\columnwidth]{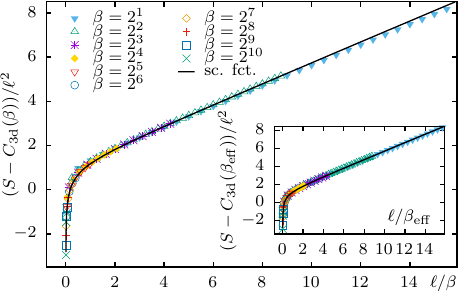}
	\caption{\label{fig:critical3dHalf} GCE subsystem entropies for the critical 3d model \eqref{eq:Htightbind} at half filling ($\mu=0$) with system size $2048^3$, and various temperatures $\beta^{-1}$. The data collapses onto the analytical scaling function \eqref{eq:Shigher-d}. Deviations at large $\ell/\beta$, due to the finite bandwidth, can be removed by using the effective temperature $\beta^{-1}_\eff$ as shown in the inset.}
\end{figure}

\section{Entanglement crossover for 3d fermions at half filling}\label{sec:3dHalfFill}
For the critical 3d tight-binding model \eqref{eq:Htightbind} at half filling, the Fermi velocity $v_\vk$ is nonzero everywhere. The prediction \eqref{eq:Shigher-d} for the crossover scaling function is hence applicable. Figure~\ref{fig:critical3dHalf} shows subsystem entropies for the GCE at various temperatures $\beta^{-1}$ and subsystem sizes, where subsystems are cubes of side length $\ell$. After subtraction of the subleading term \eqref{eq:Cd}, the data, plotted with respect to $\ell/\beta$, collapses onto the scaling function \eqref{eq:Shigher-d}. Deviations at larger $\ell/\beta$ are due to the nonlinearity of the dispersion and are again removed when using the effective inverse temperature $\beta_\eff$ instead of $\beta$ as shown in the inset of Fig.~\ref{fig:critical3dHalf}.

\bibliographystyle{prsty.tb.title}

\end{document}